\documentclass[conference]{IEEEtran}
\IEEEoverridecommandlockouts
\usepackage{cite}
\usepackage{amsmath,amssymb,amsfonts}
\usepackage{algorithmic}
\usepackage{graphicx}
\usepackage{textcomp}
\usepackage{xcolor}
\usepackage[english]{babel}
\def\BibTeX{{\rm B\kern-.05em{\sc i\kern-.025em b}\kern-.08em
    T\kern-.1667em\lower.7ex\hbox{E}\kern-.125emX}}
\begin{document}

\title{Analyzing Current Interference Situations of Connected Devices Using Context-Information and the Centralized Broker-Approach}

\author{\IEEEauthorblockN{Daniel Lindenschmitt, Michael Karrenbauer and Hans D. Schotten}
	\IEEEauthorblockA{Department of Wireless Communication and Navigation\\
		Technical University of Kaiserslautern, Germany\\
		Email:  $\lbrace$lindenschmitt,karrenbauer,schotten$\rbrace$@eit.uni-kl.de}
}

\maketitle

\begin{abstract}
The digitalization of manufacturing processes is leading to a highly increased amount of connected devices. In the course of this development a process was developed and implemented, which optimizes IEEE 802.11-systems relating to the interference situation by using context-
information. This was realized by division into two fields. First of all, data providers calculate their own interference situation to determine the optimal frequency out of the ascertained data. In this case optimal means that the chosen channel should have the lowest interference power. In a second step these providers also calculating the interference information about all Service Set Identifier (SSID) in range. The measured data is then divided into the allowed radio channel. Afterwards the gathered information will be transferred to a central space. An already existing infrastructure is used as a initial point for the further process. As soon as the provider has transferred new data, a broker can inform connected users that there is an update in the inference situation. With this developed architecture an approach to face the hidden node problem is given. Additional a system for safety-critical information is implemented.
\end{abstract}


\section{Introduction}
Digitalization has become one of the most important topics in industry. Nowadays manufacturing processes, factory halls or crafting businesses are facing the fourth industrial revolution and thus also new requirements \cite{[1]}. By looking for solutions one approach is the so called internet of things (IoT), where production systems forming an autonomic, self-controlling and sensor-based cyber-physical system (CPS). Another requirement of the new, smart production site is the wireless transmission of data. A successful transmission must be ensured especially for time- or safety-critical information \cite{[2]}. 

This paper provides an approach for a wireless transmission of data by using context-information \cite{[3]}. The developed algorithm is determining an optimal radio channel regarding to its expected interference situation, thus critical information can be transmitted with the highest probability of success. This approach can also be used to face the hidden node problem, where a node is connected to an access-point (AP) and can only communicate with this AP. This circumstance can lead to a bad interference situation in a wireless network because only the access-point calculates the Signal-to-Noise-Ratio (SNR) while the spatially distant providers may have different values for the SNR of alternative radio links. By following the deployed algorithm the SNR in a network is increased and the situation on the medium access control layer is improved.

The paper is organized as followed: In Section II the approach of the centralized broker architecture and the fundamentals of Context-Information is described. Section III explains the implementation of this architecture into the setting of a production site, while Section IV gives a detailed view of how the developed method is built up. In Section V the results of the performance analysis of the implemented method are summarized. A conclusion is drawn in chapter VI.
\section{Broker Architecture-Concept} \label{2}
In this paper the concept of a broker is used as the fundamental architecture for optimizing the interference situations in IEEE 802.11-systems. In our approach, we are using this Centralized Broker to gather all context-information which are necessary to calculate the quality of a wireless link so that the best channel in respect to the SNR can be determined. Before giving a detailed view on how context-sensitive systems are specified and organized, the term of context-information needs to be defined. After that the we will present the broker architecture that is used in this approach.

\subsection{Context-information} \label{2.1}
Context-information are already used in the nineties and since then their definition has changed over time. Early papers on context-information like the \textit{active map service} had been published \cite{[4]}. In this paper, information about the location are displayed and will be updated when a user changes his current location. Common for all this papers is, that they are using the same attributes like location, environment, identity or time to describe context-information. This is also represented in the definition from Abowd et al, where context is any information which can be used to characterize a situation of an entity \cite{[3]}. In that definition an entity is a person, a place or an object, which is relevant for the interaction between a user and an application. 

\subsection{Context-sensitive systems}
By setting up an infrastructure, we must adapt the definition for context-information in \ref{2.1}. From the view of an architecture or a system this means, that it is only achieving context-sensitivity when context is used to provide relevant information for a user. As proposed by Chen et al., there are different ways of how users can access this context-data \cite{[5]}:
\break
\begin{itemize}
	\item User access data e.g. of a sensor directly; there the reusability of a sensor is very limited
	\item Usage of a service infrastructure, where a layer infrastructure sums up the data on a low layer; this concept is suitable for reuse and the integration of new sensors
	\item Data is stored in a context-server, sensors can be reused easily and the user of a system is only responsible for querying data but not for gathering
\end{itemize}

In this approach a context-sensitive system is used, which is based on the concept of a context-server as mentioned above. That Centralized Broker is one of this context-sensitive systems, where the management of the context-information is its core task, this includes the gathering, merging and distributing of data in a system.

The implemented architecture is using a central context-server for storing and distributing the data. These data were provided by a network access. Thus the context-sensitive system is able to work efficiently, the acquisition of data need to be structured, so that only relevant information were send to the context-server. All sources of data must follow the same communication standard, so that an exchange of data is possible. Strang et al. describes different ways of how this structured communication can be achieved \cite{[6]}: 
\break
\begin{itemize}
	\item Key-value model: information will be assigned to a unique key; easy to manage but inefficient for a higher amount of data
	\item Object-oriented model: aspects like reuse and summation of data are realized via a defined interface
	\item Logic-based model: descriptive logic and logical functions are used to manage the data; an integration in an existing infrastructure is complex
	\item Markup-scheme model: usage of a hierarchical data structure with markup tags, which is based on the eXtensible markup language (XML); model is suitable for bigger networks when the XML-realization is designed efficiently.
\end{itemize}

The communication in the used Centralized Broker is based on a Markup-scheme, which is using a hierarchical data structure with eXtensible markup language. In ContextML by Knappmeyer at al. an approach for a context-sensitive system with XML is given \cite{[7]}. This approach follows the rules of a service-oriented architecture, which combines XML data with ordinary hypertext markup language (HTML) messages. 
\renewcommand{\figurename}{Figure}
\begin{figure}[h!]
	\includegraphics[scale=0.34]{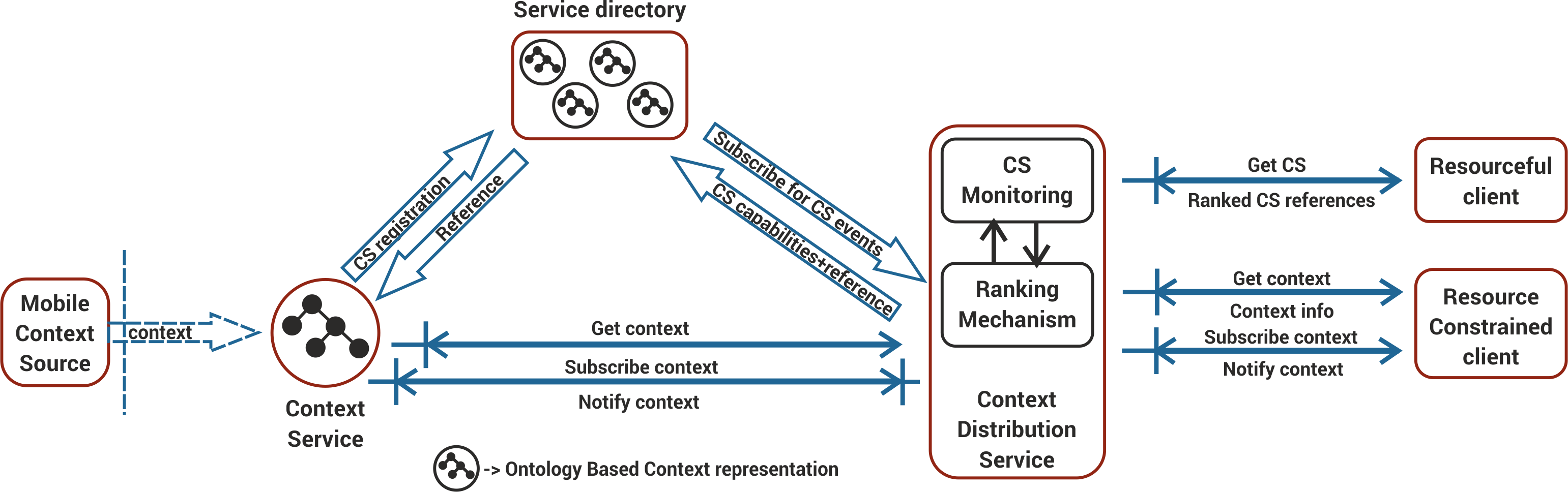}
	\caption{Context Distribution Framework \cite{[8]}}
	\label{cdf}
\end{figure}
Thus a receiver is able to get new data when available, a distribution system is necessary. One approach is the context distribution framework from Pawar et al. also shown in Figure \ref{cdf} \cite{[8]}. Here so called context-services are used. They receive the collected data directly from a source, process them and make them available to other components in that context-sensitive system. 

\subsection{Centralized Broker Architecture} \label{2.3}
In this paper we use an updated version of the ContextML approach for the centralized broker architecture \cite{[9]}. In order to determine the optimal interference situation for all connected devices it is necessary to store context-information about the localization, the safety relevance  and especially the quality of the current radio channel and the interference power of that device. 
\break
\begin{figure}[h!]
	\includegraphics[scale=0.51]{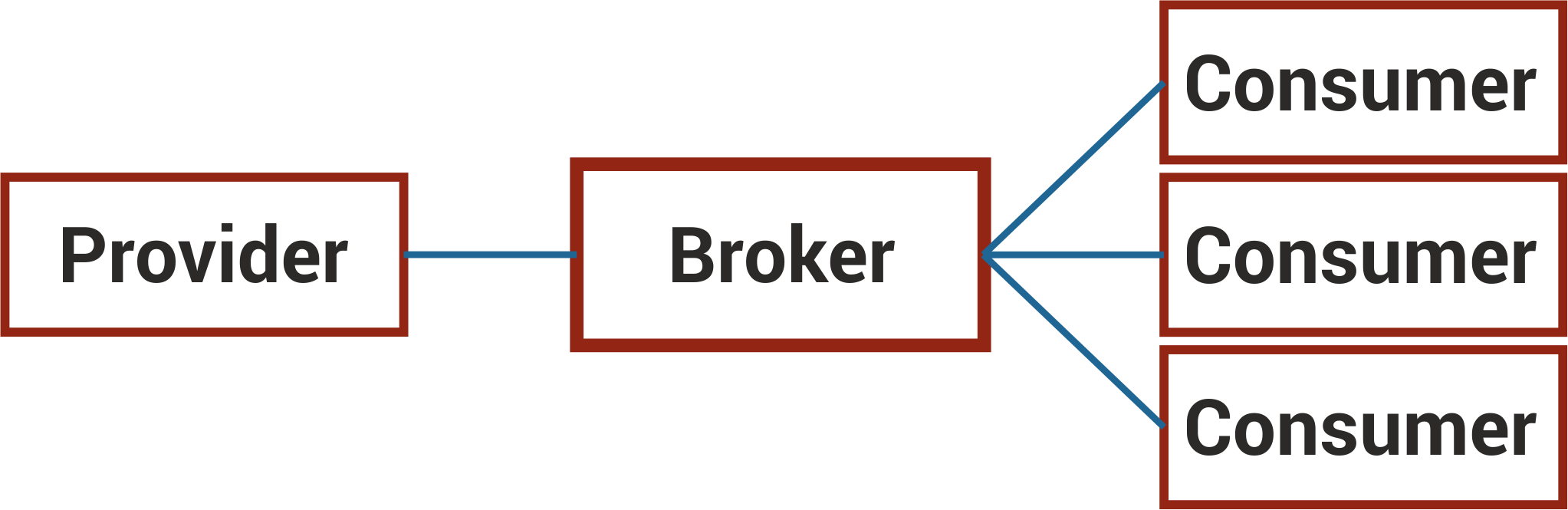}
	\caption{1-to-N network \cite{[10]}}
	\label{1ton}
\end{figure}

That so called provider delivers context-information via a network infrastructure to a central server (called broker), where the information is stored. The providers and the broker agree about permitted data types to ensure high performance within that network. The broker is used, because it can handle data within a network with a high number of sensors more efficient then systems without a centralized component. Another important aspect is that the complex task of distributing data is shifted from a simple sensor to the broker, so that a 1-to-N is created (shown in Figure \ref{1ton}). By realizing this setup it is possible that a provider transmits its data to a broker and that broker can distribute these updates to multiple user (so called consumer).

In order that all components within a network can exchange data it is necessary, that a structured communication takes place. Therefore a so called entity is defined, which has a type and an identifier. This identifier specifies a certain value out of several entities with the same type. Vice versa a type is a value which describes the summarization of entities. Besides entities we need so called scopes. In this context a scope represents a group of similar context-parameters. Every context-parameter can only be assigned to one scope. Every parameter of a scope is queried, updated or stored with all other parameters of that scope. Additional a scope has two timestamps, one shows the time of creating the data and the second gives the time, when the data will be invalid. In this approach a valid dataset can look like that:
\break
\begin{itemize}
	\item Entity identifier: noisesensor1
	\item Entity type: sensor
	\subitem Scope1: interference power
	\subsubitem ParameterA: channel 1
	\subsubitem ParameterB: channel 5
	\subsubitem ParameterC: channel 9
	\subitem Scope2: localization
	\subsubitem ParameterD: position (x/y)
\end{itemize}
In our approach the broker will ensure that there is only one valid entry per entity-scope pair, outdated data must be deleted directly. 
\section{Implementation of the context-sensitive system}\label{3}
In \ref{2} the concept of a broker-architecture was described. This will now be used to realize a context-sensitive system that is able to acquire the necessary data for calculations on the interference situation on each provider in a proper time, so that the calculated data can be used on a possible consumer before it is outdated. Therefore the detailed setup will be explained in \ref{3.1}. Afterwards the usage of the implemented system in the field of frequency management will be discussed.
\subsection{Components of the implemented system} \label{3.1}
\begin{figure}[h!]
	\includegraphics[scale=0.65]{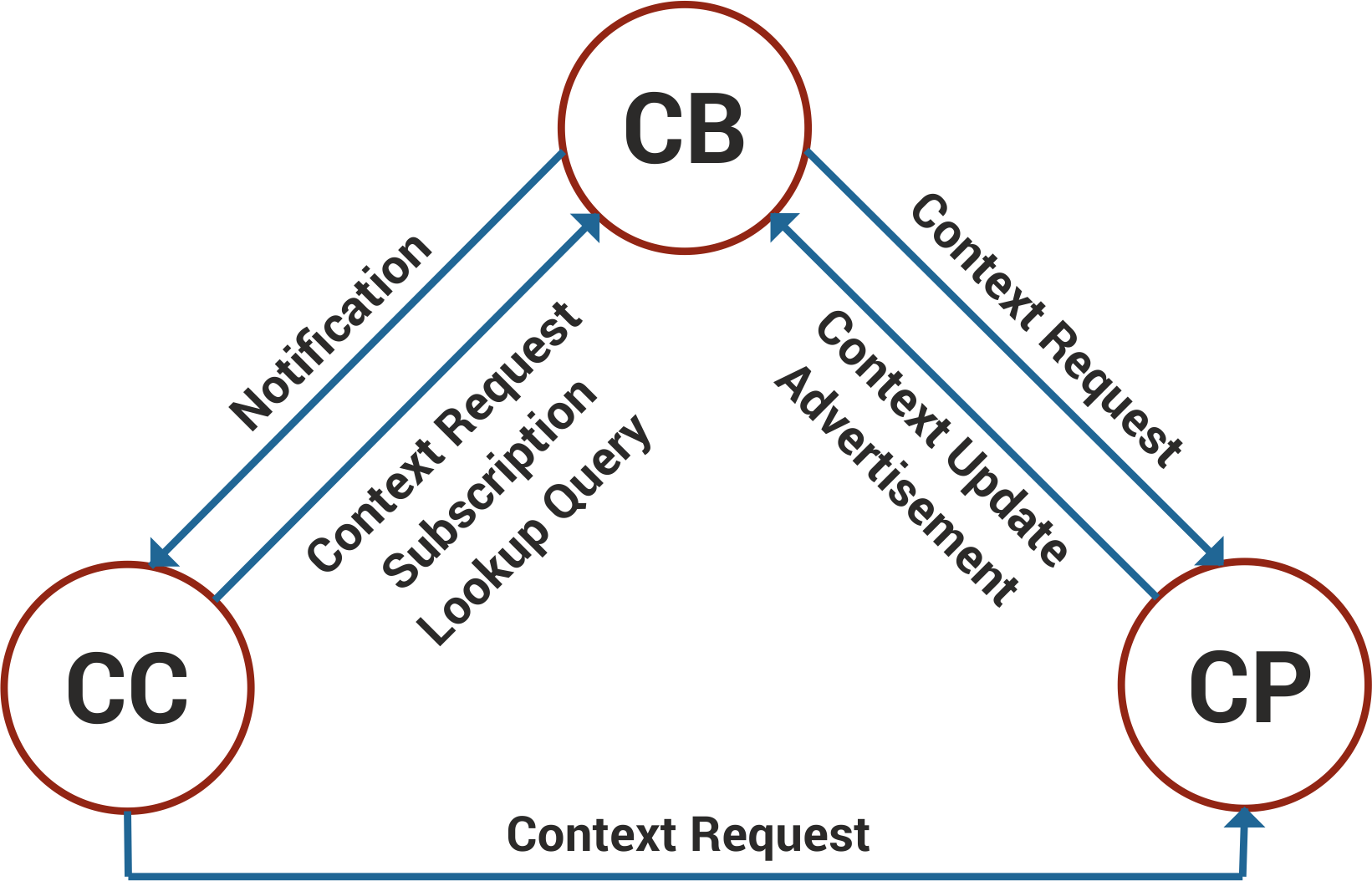}
	\caption{Communication in context-sensitive systems\cite{[9]}}
	\label{cba}
\end{figure}

The general functionality of the centralized broker architecture was explained in \ref{2.3}. In this section the detailed implementation will be explained. The approach is based on the draft by Moltchanov et al. \cite{[11]}, where three components are defined. This will be adapted and integrated to be able to perform the interference analysis in an efficient way. In Figure \ref{cba} the used system is shown.
\break
\begin{itemize}
	\item Context-Consumer (CC): In this system, the CC forms the user who is interested in context data. Therefore it has to be able to receive any relevant data. He must specify the entity type, identifier and scope. 	
	\item Context-Provider (CP): In the system, the CP is responsible for making context data available, which it receives e.g. from sensors. These must be processed in such a way that they are available in the used schema. The following functions for the CP are available:
	\subitem Advertisement: The first task of a CP is to register with a CB. He informs the broker about his entity and scopes as well as the parameters.
	\subitem Context-Update: Every CP registered in the CB must provide context data in a certain frequency to ensure that the data in the CB is always up-to-date.
	\subitem Direct Request: If the CC sends a request to the CP, it must be able to send data to the CC.
	\item Context-Broker (CB): The CB represents the central component in the architecture because it regulates the flow of information between source and user.	Therefore the following functions can be used:
	\subitem Provider Registry: The CB maintains a list of all registered CPs including their entities and scopes. CC can query this list.
	\subitem Cache: All current data of the providers registered on the broker are stored in the cache. If a CC now requests data, the current data is first searched for in the cache. If these are not available, the CB sends a request to the CP.
	\subitem Subscription: Each CC can subscribe to a specific entity-scope pairing. If the broker receives new data from a subscribed entity-scope pairing, it will automatically check the subscribers and inform them if necessary.
\end{itemize}
The communication between the context provider and the context broker will be relevant for the further course of the work. More detailed information about communication as well as increased efficiency through decentralized solutions can be found in \cite{[9]} and \cite{[10]}.
\subsection{Usage in frequency management} \label{3.2}
In this section the usage of the explained architecture with its three components will be combined with the task of determing an optimal radio channel in respect to the interference situation. 

As Context-Provider a device with an active radio link to a WiFi is used. This provider generates data in a defined time interval. This data contains information about the interference power for all relevant WiFi-channels in the used 2.4 GHz ISM band as well as the current location of the device and a value about the safety-critical status. The tasks of the Context-Broker were already specified in \ref{3.1}. The broker stores the data transmitted by all connected providers, which update their information in specified intervals. If the data of a provider is outdated, the broker will contact this provider to check the availability.
\section{Method of determining an optimal radio channel}
In the following section the method of determining an optimal radio channel is presented. Before the deployed protocol and determining process is explained, the necessary technical fundamentals will be summarized.
\subsection{Technical fundamentals}\label{4.1.}
As already mentioned an essential aspect of the approach is the determination of an optimal radio channel under the use of context-information. This raises the question, how this optimal radio channel is defined. To answer this question it is also necessary to deal with the topic of wireless networks, the channel selection and the data throughput.

In this approach an implementation in a wireless local area network (WLAN) in the 802.11n standard at 2.4 GHz is chosen because of the advantages like ad hoc connectivity and the sufficient radio coverage in comparison to 5 GHz(up to 30 meters indoor and 300 meters outdoor with 100 mW transmission power\cite{[13]}). Even if the maximal bitrate of 2.4 GHz WLAN is lower then of a 5 GHz network, it is still sufficient for the usage in this approach. It is also possible to adapt this determining process on other non-cellular wireless techniques as e.g. bluetooth or low power wide area networks. WLAN is performing in the Industrial, Scientific and Medical (ISM)-band, therefore it is not necessary to apply for a license at the responsible authority. 
\break
\begin{figure}[h!]
	\includegraphics[scale=0.43]{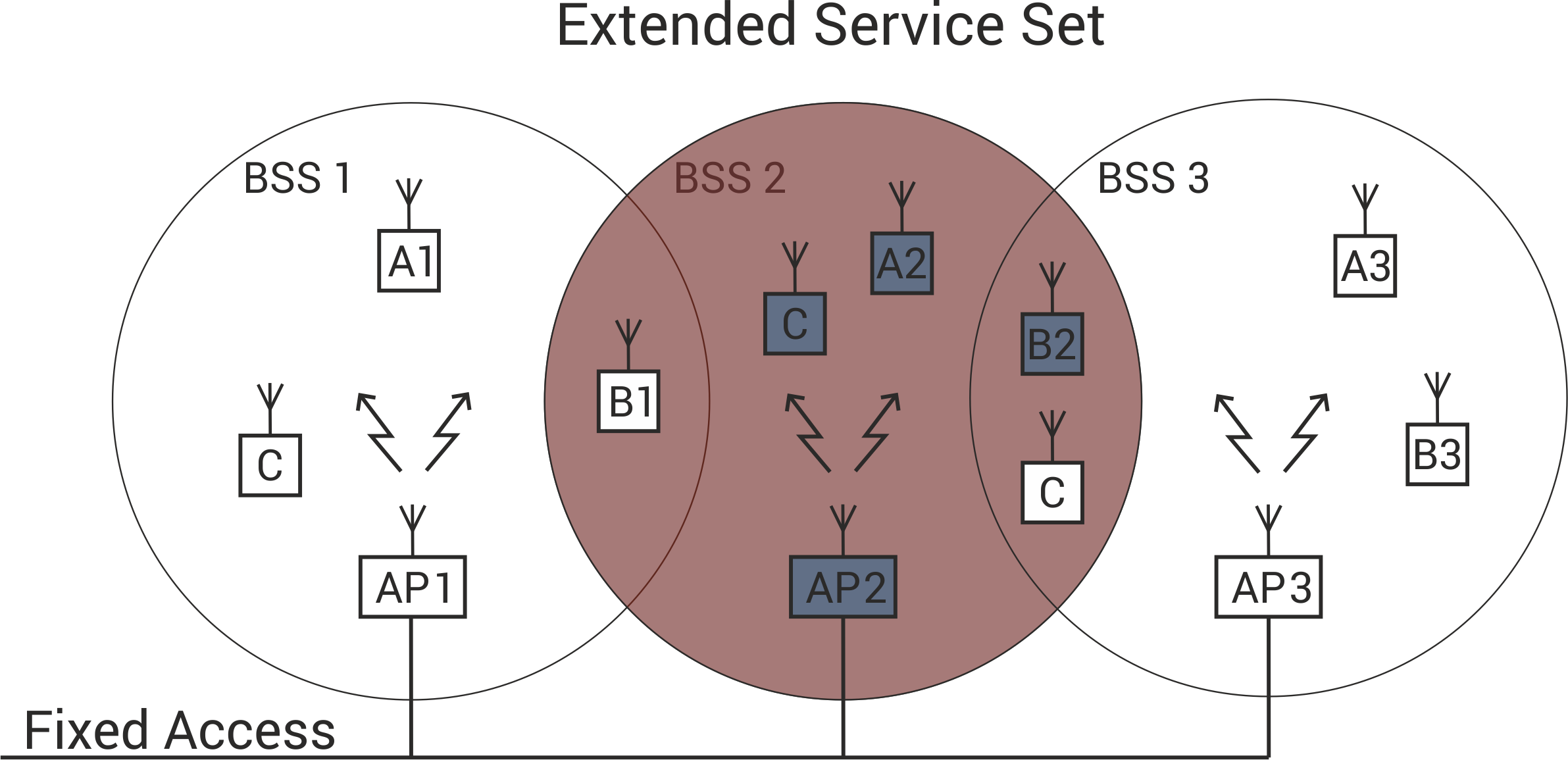}
	\caption{WLAN infrastructure with acess point\cite{[13]}}
	\label{ess}
\end{figure}

\begin{figure}[h!]
	\includegraphics[scale=0.188]{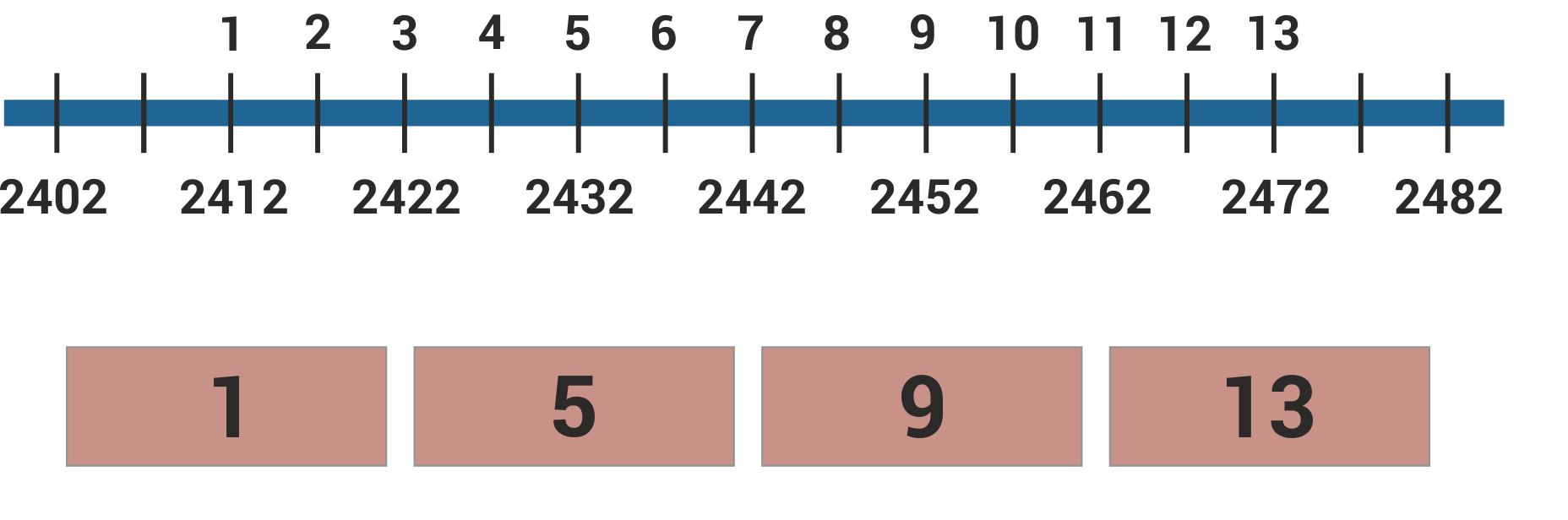}
	\caption{Carrier frequencies according to IEEE 802.1n\cite{[14]}}
	\label{kanalaufteilung}
\end{figure}

In WLAN all participants in the wireless network have to use the same frequency band and radio channel. The necessary organization like control functions of the network is done in the access point, which is also demonstrated in figure \ref{ess}. The frequency band in 2.4 GHz WLAN has 13 channels with a carrier frequency on 
\break
\begin{equation}
	f_c = 2.407MHz+(5MHz*n) \text{ für } n= 1 ... 13\label{fc}
\end{equation}
As specified in IEEE 802.11n a bandwidth per channel of 20 MHz is used \cite{[14]}. To ensure that the used radio channel is not interfering close-by networks (as shown in figure \ref{ess}) the usage of four non-overlapping carrier frequencies are defined, which is also illustrated in figure \ref{kanalaufteilung}.

In that case one is considered as optimal if each transmitted signal can be received correctly. The channel which comes closest to this value should be selected for the respective transmission, since correct reception is thus given with the highest probability. The signal-to-noise ratio (SNR) indicates the quality of a wireless link and can be calculated as follows
\break
\begin{equation}
	SNR=\dfrac{signal power}{noise power}=\frac{P_{signal}}{P_{noise}}\label{snr1}
\end{equation}
\break
\begin{equation}
	SNR_{F1}=\dfrac{G_{t}*G_{r}*P_{t}}{k*T*B*F}\label{snr2}
\end{equation}
with $G_{t}$ und $G_{r}$ as antenna gain, $P_{t}$ as signal power (here 100 mW) and $k$ as Boltzmann constant, the noise temperature $T$, the bandwith $B$ and the absorption of free-field transmission $F$. The assumption that there is no signal gain through the used antennas leads to the following equation
\begin{equation}
	SNR_{F1}\approx\dfrac{0,100W}{8,085*10^{-8}}\approx 90dB\label{snr3}
\end{equation}
This represents the optimal case of a signal-to-noise ratio. In realistic applications, however, additional interferences like other radio transmitters in the same network or from neighboring networks or disruptions due to multipath propagation can occur. This means that the formula for the SNR has to be adjusted as follows
\begin{equation}
	SNR_{F2}=SNR_{F1}*\dfrac{1}{I_{M}*I_{N}}\label{snr4}
\end{equation}
with $I_{M}$ for disruption due to multipath propagation and $I_{N}$ for disruption from additional participants in the same network or in neighboring areas.
\subsection{Deployment of the protocol}\label{4.2}
In the previous subsection the necessary fundamentals were introduced. Especially the definition of the optimal radio channel (as defined in equation \ref{snr4}) and the chosen radio transmission technique is important for the integration of the deployed protocol and determining process. To ensure a well-defined communication between all components (as mentioned in \ref{2} and \ref{3}) it is necessary to deploy a protocol for standardized data exchange. Therefore the in section \ref{3.1} introduced functions have to be revised. The integrated protocol is based on the work of Pereira \cite{[12]}, where the XML data format has been transformed into strings.
\begin{center}
	\textit{Header$\vert$(char) Flag (A) or Flag (U)$\vert$(string) Provider ID $\vert$(string) Entity type$\vert$(string) Entity ID$\vert$ (string)  Scope$\vert$(string) Timestamp begin$\vert$(string) Timestamp end$\vert$Payload}
\end{center} 

A new provider registers at the broker with an advertisement-message. Flag A indicates that this message is an advertisement, U indicates a context update. This is relevant because both messages use the same message format and therefore only differ in this flag. Furthermore, the ID of the provider, the entity type, the entity ID and the beginning and end of the validity range of the data are transmitted as a UNIX-timestamp in seconds. The actual payload, e.g. the determined context information, is then transferred. The broker responds to all messages with either a
Acknowledgment (ACK) or not-acknowledgment (NACK). Direct requests from the Context-Consumer to the Context-Provider have been eliminated from this approach due to performance reasons. A communication now only takes place between Context-Provider and Context-Broker as between Context-Consumer and Context-Broker. 

While the mentioned message types represent the framework for standardized and thus efficient communication, the actual payload contains the respective context information of a provider. This payload follows the scheme
\break
\begin{center}
	\textit{Security channel/Channel recommendation/Channel switch/Interference power/Position X/Position Y}. 
\end{center}
The security channel contains information on whether the provider is a critical system whose data transmission must primarily be successful or not. The security channel can assume the value 0 for not critical or 1 for critical. The channel recommendation contains the optimal radio channel determined by the provider with respect to the lowest interference power and specifies the value as a frequency in MHz. The channel switch indicates whether a change of the radio channel is necessary from the provider's point of view, since a channel with better interference properties is available. It can take the value 0 for no change or 1 for change. The interference power indicates the current value of the interference power in dBm. Finally, the data for the localization of the provider is transferred.
\subsection{Deployment of the determining process}
As the last step in implementing this approach the deployment of the determining process takes place. Here the core task is the generation of context-sensitive values for the interference situation at the respective context-providers. It is then necessary that the measured values are calculated at the provider and transmitted to the broker at fixed intervals afterwards.
In order to deploy a well-suited process it is necessary to answer the question under which conditions a certain channel is selected as a recommendation and suggested to the access point.

At first the definition of basic settings takes place, so that communication between the individual components is possible. After defining general values, the actual program is executed by calling the Main()-method. The first step in the execution is to initialize the system. While initializing, the necessary settings for operating the provider are loaded from a configuration file. By relocating the settings to a separate text file, providers can be put into operation more quickly on site, as no recompilation is necessary. Furthermore, these can be adapted more quickly to changed circumstances. 
After the initialization phase of the provider, the actual execution of the program starts. A distinction is made between a security-critical or non-security-critical application. Providers with this mark also have the selected security channel available to determine an optimal radio channel. However, this does not necessarily always has to use the corresponding security channel if one of the other three radio channels can have better interference properties. 

After the provider goes into analysis mode, a ping is first sent to the broker to ensure that it is available. If the ping request fails, the provider issues a message and ends the execution. If the ping is properly confirmed by the broker, the provider begins with the interference analysis. First of all, there is an initial radio check. The aim of this initial check is to be able to assign one of the four defined radio channels to each provider as quickly as possible so the listening mode can be started. In this mode the continuous analysis of the inference performance is performed. A period of validity for the measured values is specified in the configuration file. This is adapted in the listening loop. If the provider is in a radio network that rarely needs a channel change, the period of validity increases and it decreases if a channel change is more often necessary. As part of the interference measurement, the connection currently being used is first analyzed and the frequency and dBm value noted. This dBm value corresponds to the Received Signal Strength Indicator (RSSI). The RSSI value already takes into account e.g. the antenna reinforcement or free space attenuation. It is important to mention that the RSSI value does not follow an exact definition and is therefore interpreted differently by different hardware. In the implemented procedure, the provider scans all available SSIDs and saves the respective frequency and the dBm value.

A radio channel is used when its averaged interference power is the lowest, since in this case it is assumed that potential SSIDs with this frequency e.g. have a bigger distance to the own access point or a lower transmission power. By changing the frequency of the access point, the best SNR value can be achieved and the channel is then defined as optimal. The collected data are then filtered so that only frequencies according to the optimal distribution are considered. 

The mean value of the signal power is then calculated for each radio channel, the measured value from the currently used SSID while be filtered so that the currently used radio link does not play a role in the assessment of the interference power. This is necessary because otherwise the measured values will be falsified. If the provider determines that a change of channel is necessary, because one of the channels has a lower interference power, it sets the channel switch flag to 1. The message is then transmitted with the determined values. The presented method enables a continuous selection of an optimal radio channel determined at the provider.

\section{Performance analysis}\label{5}
After the deployed algorithm for analyzing the interference situation in order to face e.g. the hidden node problem has been described, now a analysis of the performance of this approach takes place.

\begin{figure}[h!]
	\includegraphics[scale=0.61]{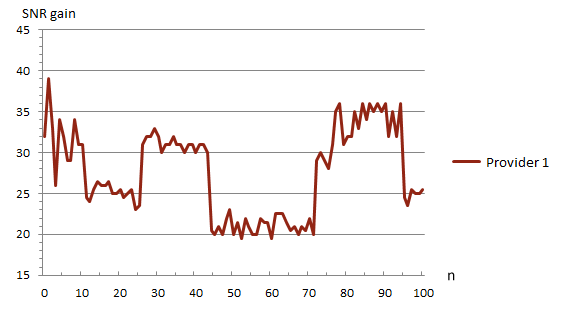}
	\caption{Performance of a provider with safety-critical data}
	\label{chanel selection}
\end{figure}

The main task is the determination of the best channel according to the lowest interference power at the provider. As part of the provider tests, 100 measured values were recorded for each analysis. In the test scenario the provider recommends channel 9 in the measured values, as no SSID radio transmissions are carried out on this. This results in a SNR gain shown in figure \ref{chanel selection}, since the current connection is also located on channel 13. The displayed SNR gain can be achieved by changing the current used radio channel which is indicated by the switch flag in the protocol explained in \ref{4.2}.
\break

As figure \ref{chanel selection} shows there are multiple sharp decreases of the SNR gain e.g. at measurement 12 or 44. The reason for that behavior of the provider 1 is due to a change in channel recommendation, which can have different occasions like a temporary disruption on the radio channels or change in the radio environment of the testbed. In this case the provider 1 changes his current channel recommendation to a radio channel with a lower interference power. To prevent that the provider goes into a deadlock an avoiding system should be implemented in a further work.
\section{Conclusion}
This paper provided an approach for a wireless transmission of data by using context-information \cite{[3]}. The developed algorithm is determining an optimal radio channel regarding to his expected interference situation, thus critical information can be transmitted with the highest probability of success. This approach can also be used to face the hidden node problem, where a node is connected to an access-point and only can communicate with this AP. The deployed algorithm is based on a centralized broker architecture introduced in \ref{2} where multiple so called providers can send data to a central point. This architecture was adapted in\ref{3} to fit into the system of using context-information for analyzing the radio link quality. The revised provider are responsible for continuous selection of an optimal radio channel determined. As shown in section {5} the deployed algorithm can increase the SNR in a wireless network and the situation on the medium access control layer is improved. For following the integration of a context-consumer can be realized so a better overview about the wireless link situation at the access point is achieved. 

\section*{Acknowledgment}
This work has been supported by the Federal Ministry of Education and Research of the Federal Republic of Germany (Foerderkennzeichen 16KIS0725K, 5Gang). The authors alone are responsible for the content of the paper.

\end{document}